\documentclass[9pt,twocolumn,twoside]{osajnl}
\journal{ol} % Choose journal (ao, aop, josaa, josab, ol, pr)
\newcommand{\avg}[1]{\left\langle #1\right\rangle}
\newcommand{\de}[0] {{\rm d}}
\newcommand{\eeqref}[1]{Eq.~(\ref{#1})}
\newcommand{\tr}[0]{{\rm Tr}}

\setboolean{shortarticle}{true}
\usepackage{xcolor}
\begin{document}
	
	\title{Measuring fluorescence into a nanofiber\\by observing field quadrature noise}
	
	\author[1]{Shreyas Jalnapurkar}
	\author[1]{Paul Anderson}
	\author[1]{E. S. Moiseev}
	\author[1]{Pantita Palittapongarnpim}
	\author[2]{Andal Narayanan}
	\author[1]{P. E. Barclay}
	\author[1,3,4,*]{A. I. Lvovsky}
	
	\affil[1]{Institute for Quantum Science and Technology, University of Calgary, Calgary AB T2N 1N4, Canada}
	\affil[2]{Light and Matter Physics Group, Raman Research Institute, Bangalore 560080, India}
	\affil[3]{Department of Physics, University of Oxford, Oxford OX1 3PU, UK}
	\affil[4]{International Center for Quantum Optics and Quantum Technologies (Russian Quantum Center), Skolkovo, Moscow 143025, Russia}
	
	\affil[*]{Corresponding author: Alex.Lvovsky@physics.ox.ac.uk}
	
	%% To be edited by editor
	% \dates{Compiled \today}
	
	\ociscodes{(270.5290)   Photon statistics; (270.5570)   Quantum detectors; (060.2920)   Homodyning; (060.2430)   Fibers, single-mode.}
	
	%% To be edited by editor
	% \doi{\url{http://dx.doi.org/10.1364/XX.XX.XXXXXX}}
	
	\begin{abstract}
		We perform balanced homodyne detection of the electromagnenic field in a single-mode tapered optical nanofiber surrounded by rubidium atoms in a magneto-optical trap. Resonant fluorescence of atoms into the nanofiber mode manifests itself as increased quantum noise of the field quadratures. The autocorrelation function of the homodyne detector's output photocurrent exhibits exponential fall-off with a decay time constant of $26.3\pm 0.6$ ns, which is consistent with the theoretical expectation under our experimental conditions. To our knowledge, this is the first experiment in which fluorescence into a tapered optical nanofiber has been observed and measured by balanced optical homodyne detection.  
	\end{abstract}
	
	\setboolean{displaycopyright}{true}

	\maketitle

	\section{Introduction}
	A tapered optical nanofiber  (TNF) \cite{Birks1992} is manufactured by heating and elongating a regular optical fiber, reducing its diameter below the optical wavelength. A significant fraction of the optical mode guided by such a fiber propagates as an evanescent field, which can interact with atoms or artificial atom-like objects placed in the neighborhood of the fiber. Because the field stays focused over a macroscopic length determined by the profile of the TNF, it can interact with a large number of such objects \cite{Morrissey2013,Nieddu2016}. This makes TNF an attractive tool for various quantum light-matter interfacing applications, such as quantum-optical memory \cite{Sayrin2015,Gouraud2015} and cavity quantum electrodynamics \cite{Keloth2017}. 
	
	The TNF has the property to ``suck in" spontaneous emission from atoms in its direct neighborhood: due to Purcell-like effects, a significant fraction of that emission occurs via the guided mode \cite{LeKien2005,Nayak2007}. This effect has been utilized, for example, for measuring the temperature of magneto-optical traps \cite{Russell2013,Russell2014,Grover2015} as well as sensing position of atoms in optical lattices \cite{Grover2015}. It is also an excellent tool to study the undesired process of atom's condensation on the nanofiber's surface \cite{Nayak2012}. Resonant fluorescence, an inherently multimode phenomenon, can be detectable in a single mode of the TNF, even if produced by only a few emitters. This makes fluorescence amenable to homo- and heterodyne methods, which can detect light only in the optical mode of the local oscillator. Detection of fluorescence by balanced homo- and heterodyne techniques has been performed for an atom \cite{squeezed_light_from_atom}, trapped ion \cite{singleionfluorescence} and driven superconducting artificial atom \cite{rft} in a high finesse cavity, as well as for the observation of the motion of multiple atoms in magneto-optical traps \cite{Westbrook1990,Jessen1992}. 
	%Since the field from a few atoms is very weak, the detection technique obviates the need for single-photon detectors. Although fluorescence has been detected in previous experiments by single-photon detectors, balanced homodyne detection is a powerful tool capable of completely characterizing any arbitrary quantum state of light.
	
	\begin{figure}[htbp]
		\centering
		\includegraphics[width=0.9\linewidth]{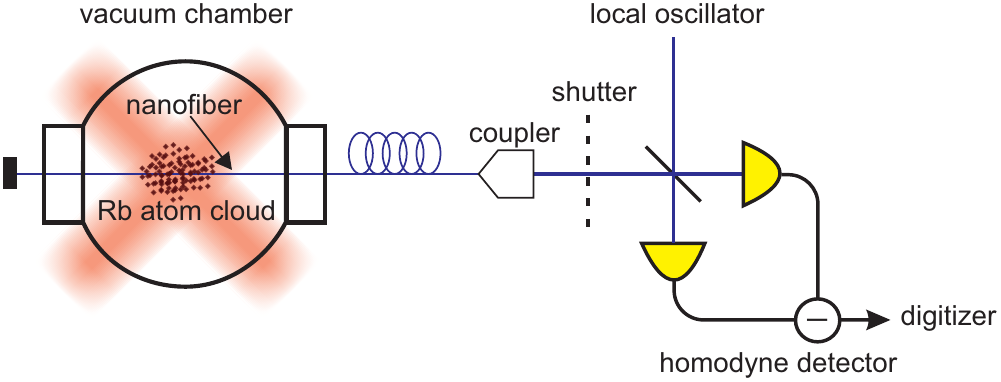}
		\caption{Schematic diagram of the experiment. The cold atomic cloud is overlapped over the TNF in the vacuum chamber. The shutter alternates the acquisition of the vacuum field and the signal from atoms by the BHD.}
		\label{schematic}
	\end{figure}
	
	Here we measure resonant fluorescence into a TNF by means of ``classic'' balanced homodyne detection with a strong local oscillator (Fig.~\ref{schematic}), by analyzing the noise statistics of the detector's output photocurrent in the time domain. Previously, non-balanced heterodyne detection at the single-photon level with a microscopic local oscillator has been used to measure the spectrum of  the fluorescence into the nanofiber \cite{Das2010}. This method was later applied to probe motional sidebands \cite{Meng2018} and observe ultra-strong spin-motion coupling \cite{Dareau2018} of the atoms trapped by the evanescent field surrounding the nanofiber. The advantage of balanced homodyne detection is the capability of completely characterizing an arbitrary quantum state of light \cite{Lvovsky2009}, which is particularly important in the context of quantum light-atom interfacing. Furthermore, by applying this technique to resonant fluorescence, important insights into the physics of this process can be gained \cite{beltranBHD}. %which obviates the need for single-photon detectors. %Resonant fluorescence has been detected in earlier experiments for atoms in a cavity, superconducting devices and in ions. To our knowledge this is the first experiment measuring the fluorescence rates in presence of a tapered nanofiber. Although the purcell enhancement is not a new phenomenon, %
	If, in addition, the temporal autocorrelation behavior of that photocurrent is analyzed, one can obtain the information about the temporal mode in which a given optical state (e.g.,~a single photon) is generated \cite{Macrae2012,Morin2013,Qin2015}.

	Our experiment consists in continuous measurement of the output photocurrent of the detector, which is proportional to the instantaneous measurement of the electric field quadrature of the single TNF mode. We extract the information about the fluorescence spectrum from the autocorrelation statistics of this photocurrent.  Specifically, the difference between the autocorrelation profiles in the presence and absence of rubidium atoms around the nanofiber reproduces the exponential temporal profile of the photons emitted thanks to resonant fluorescence. %It allows us to measure the fluorescent lifetime of rubidium atoms. % as well as the average number of rubidium atoms in the nanofiber's neighborhood.
	
	\section{Measuring temporal profile by homodyne detection}

	A simple way to visualize our technique is to think of the optical field  spontaneously emitted by the atoms into the fiber as a stationary stochastic process. The balanced homodyne detector (BHD)'s output voltage $V(t)$ is proportional to the amplitude of that field in the fiber mode as a function of time; hence, the autocorrelation $\avg{V(t)V(t+\Delta t))}_t$ can be calculated. According to the Wiener-Khinchin theorem, the Fourier transform of this autocorrelation is equal to the power spectrum of the resonant fluorescence, which is a Lorentzian line whose width, $\gamma$, is evaluated below. The autocorrelation function can be expected to be double-exponential with a decay constant of $2/\gamma$.
	
	However, the measured signal is microscopic, and hence contaminated  by the shot noise of the local oscillator. In order to quantitatively predict the spectrum and amplitude of the signal, we apply the quantum analysis of Refs.~\cite{Macrae2012,Morin2013,Qin2015}. In this analysis, time is treated in terms of short discrete bins of duration $\tau$ in order to accommodate the discrete nature of digital data acquisition: the $k$th bin corresponds to the moment $t=k\tau$. Suppose photons with the same temporal density matrix  $\rho_{jk}$ are repeatedly emitted into a certain spatial mode. The time-dependent electromagnetic field quadrature in this mode is being measured with a BHD with the quantum efficiency $\eta$. The mean autocorrelation of the quadrature values at time bins $j$ and $k$ is then given by  \cite{Macrae2012,Morin2013,Qin2015}
	\begin{equation}\label{autocorr1}
	\avg{X_jX_k}=\frac12\delta_{jk}+\eta{\rm Re}\rho_{jk}.
	\end{equation}
	Here the first term corresponds to the shot noise (the vacuum field whose quadrature variance is normalized to $\avg {X^2}=\frac12$) and the second term to the field of the photon \cite{Qin2015}. In the case of spontaneous emission, \begin{equation}\label{rhospont}
	\rho_{jk}=\theta(j)\theta(k)\gamma\tau e^{-\frac{\gamma\tau}2(j+k)},
	\end{equation}where $\theta(\cdot)$ is the Heaviside step function and the factor of $\gamma\tau$ is due to normalization, so that $\tr\rho_{jk}=1$ for $\gamma\tau\ll1$.
	
	Equation (\ref{autocorr1}) assumes that the photons are repeatedly emitted at the same time with respect to a certain reference. In our case, the emission occurs at random times, so the autocorrelation has to be averaged over random moments $m$ of the emission
	\begin{equation}\label{autocorr2}
	\avg{X_jX_k}=\frac12\delta_{jk}+\eta{\rm Re}\avg{\rho_{j-m,k-m}}_m.
	\end{equation}
	If $f\ll\gamma$ is the average frequency at which photons are emitted, $m$ can be assumed to be uniformly distributed in the interval from $-1/(2f\tau)$ to $1/(2f\tau)$.
	Substituting the density matrix (\ref{rhospont}), we find $\avg{X_jX_k}=\frac12\delta_{jk}+\eta f\tau e^{-\frac{\gamma\tau}2|j-k|}$, which can be rewritten as 
	\begin{equation}\label{autocorr3}
	A_X(l)\equiv\avg{X_jX_{j+l}}_j=\frac12\delta_{l,0}+\eta f\tau e^{-\frac{\gamma\tau}2|l|}.
	\end{equation}
	Remarkably, the quadrature variance ($A_X(0)$) does not depend on the spontaneous emission rate $\gamma$. This is because the pulse associated with each photon (whose height is proportional to $\gamma$, and width is inversely proportional to $\gamma$) is randomly distributed over a time interval that is much larger than its duration. 
	
	It remains to account for the electronic noise and bandwidth of the BHD. The role of the electronic noise is equivalent to an optical loss of about 10\% \cite{Hoffman}, and therefore insignificant for our treatment. The effect of the final bandwidth is to relate the BHD output voltage  to the field quadratures by the convolution $V_{j}=\sum_{j'} X_{j'} R_{j-j'}$, where $R$ is the detector's temporal response function. Accordingly, the  autocorrelation profile of the voltage is related to that of the quadratures via the convolution
	\begin{equation}\label{autocorr4}
	A_V(l)\equiv\avg{V_jV_{j+l}}_j=\sum_{l'}A_X(l')A_{V0}(l-l'),
	\end{equation}
	where $A_{V0}(l-l')=\sum_iR_iR_{i+l-l'}$ is the convolution of the response function with itself. This function is equal to the autocorrelation exhibited by the BHD output voltage in response to the vacuum field, in which case only the first term in \eeqref{autocorr3} is present, and can therefore be easily measured  [Fig.~\ref{fig:result}(a)]. The difference between the voltage  autocorrelation profiles in response to the atomic fluorescence and the vacuum field yields the second term in \eeqref{autocorr3} which is of interest to us. Although this second term is also distorted by the convolution with $A_{V0}$, this effect is negligible because, as evidenced by Fig.~\ref{fig:result}(a), the response time of the detector is $\sim 5$ ns, much shorter than the relevant lifetime of the $D_2$ transition in Rubidium-87 ($T_1=26$ ns) \cite{steck2001rubidium}. Therefore we can approximately write
	\begin{equation}\label{autocorr5}
	A_V(l)\approx\frac12A_{V0}(l)+\eta f\tau e^{-\frac{\gamma\tau}2|l|}.
	\end{equation}
	The first term in Eq.~(\ref{autocorr5}) provides us with natural means to find the proportionality coefficient between the experimentally observed output voltage of the BHD and the corresponding quadrature. Knowing this coefficient, we can then determine the magnitude  of the second term and hence the rate $f$ of photon emission into the nanofiber.

	\section{Experiment and results}
	The tapered nanofiber (TNF) is manufactured by pulling a commercial single mode fiber by flame brushing technique \cite{flamebrush}. This technique allows flexibility in choosing the radius and length of the waist of the TNF. For our experiment, we keep a radius of $230 \pm 20$ nm and a waist length of 10 mm. The TNF is loaded in an ultra high vacuum  chamber where Rubidium-87 atoms are prepared in a magneto optical trap (MOT). The position of the MOT is manipulated by three orthogonal compensating coils to ensure the best coupling. The TNF transmission efficiency was 90\% immediately after fabrication, but degraded to 78\% when placed into the vacuum chamber. 
	
\begin{figure}[t]
		\centering
		\includegraphics[width=0.9\linewidth]{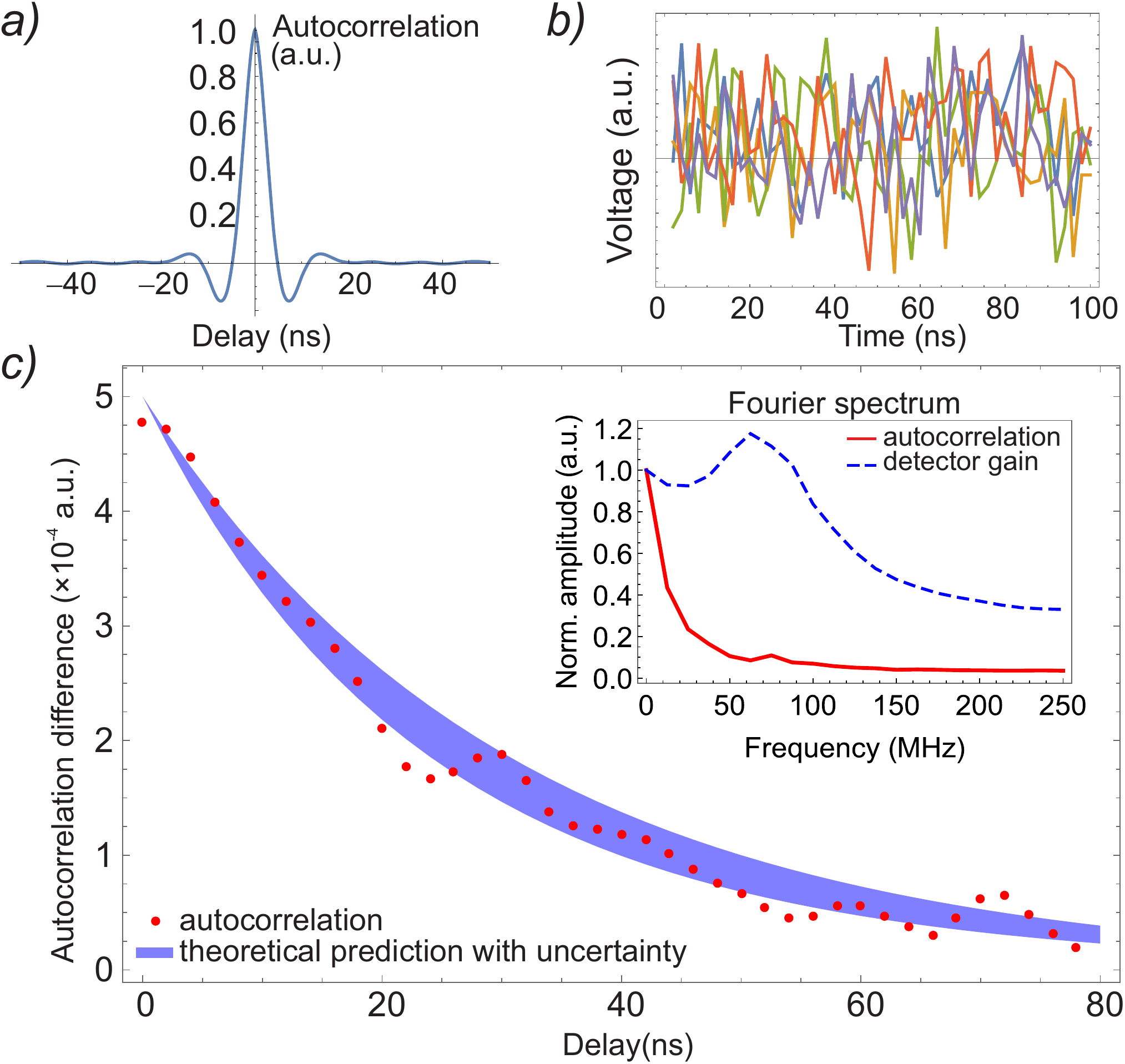}
		\caption{Experimental results. a) Auto-correlation $A_{V0}(l)$ of the signal from the BHD without atoms. b) Five samples of the BHD output signal. c) Difference $A_{V}(l)-A_{V0}(l)$ between the BHD signal autocorrelation profile in the presence and absence of atoms, corresponding to the resonance fluorescence. The uncertainty in the theoretical prediction is due to that in the fiber radius.  The red curve in the inset shows the autocorrelation spectrum, which is largely Lorentzian. The feature around $75$ MHz is due to the gain spectrum of the detector. The vertical axis units in (a) and (c) are consistent with each other.}
		\label{fig:result}
	\end{figure}

	The TNF collects fluorescence of the atoms excited by the MOT beams. We take the output signal from one end of the TNF and connect it to a 90:10 fiber beam splitter. The 10\% output is directed to a single photon counting module to monitor the photon emission rate. The other output of the beam splitter is taken as the signal for homodyne detection, which is realized by a home-made BHD \cite{Masalov2017} with the linearly polarized local oscillator tuned to the expected emission frequency. The quantum efficiency of the BHD photodiodes is 0.91, which has to be multiplied by an additional loss factor of 0.85 occurring due to imperfect overlap between the LO and signal modes. An additional loss factor of $1/2$ is present because the homodyne detector is only sensitive to the polarization mode of the local oscillator. The BHD output data are collected by an 8-bit digitizer card (Acqiris DP-240) with a resolution of $\tau=2$ ns. A few samples of the collected data are shown in Fig.~\ref{fig:result}(b).
	
	The MOT cooling laser is resonant to the $5S_{1/2},F=2\to5P_{3/2},F'=3$ cycling transition (selection rules prohibit the atom to decay into the $F=1$ ground state). Those atoms that do decay outside $5S_{1/2},F=2$ are repumped by the MOT repumping beam. Thus, majority of the photons coupled into the fiber are at the frequency of the cooling transition. The SPCM gives an estimate of the emission rate from the atomic cloud after losses in the propagation of the TNF mode. The observed photon count rate of the SPCM is $\sim 10$ KHz. The SPCM has 60\% detection efficiency at wavelength of 780 nm giving a photon emission rate of $f\sim 150$ KHz. This means that the ratio between the signal from the atoms, given by the second term in Eq.~(\ref{autocorr5}), and the shot noise, given by the first term, is on a scale of $2f\tau/2\approx3\times10^{-4}$. In the experiment, we observe this ratio to be $\sim 2.5\times 10^{-4}$. 
	
	In order to obtain a better signal to noise ratio, we acquire large sets of data. Specifically, the data corresponding to $5\times10^{10}$ 2-ns bins are acquired and analyzed. This corresponds to a relative uncertainty of $\sim5\times10^{-6}$ for each point of the autocorrelation profile.  But because the atomic signal itself is very small in comparison with the vacuum autocorrelation, this signal has a relative statistical uncertainty on a scale of 2\%, and is hence visibly noisy [Fig.~2(c)]. Additionally, the variance of the shot noise current drifts due to fluctuations in the LO power. To circumvent this issue, we alternate the acquisition of the signal from atoms and vacuum with the help of a shutter as seen in Fig.~\ref{schematic} with a period of 2 s. The difference in auto-correlation functions is calculated for alternate sets and averaged over multiple sets.

	%Ideally, the cooling laser should be used as the LO since it is also driving the atomic transitions. Since majority of the laser power is used up to generate the MOT, there was not enough power left for the LO (requires about 10 mW). A separate External cavity diode laser (ECDL) was used as the LO and tuned to the atomic frequency. The LO frequency was monitored by a separate OBHD and maintained within $\pm 3$ MHz of the cooling transition which is well within the bandwidth of the detector. These fluctuations are much slower than the expected atomic decay rate and thus do not affect the measurements.

	Figure 2(c) shows the difference of the BHD voltage autocorrelation in the presence and absence of atoms. The autocorrelation profile resembles an exponential decay with the time constant of $2/\gamma=26.3 \pm 0.6$ ns. This is about twice as fast as one would expect for a free spontaneously decaying rubidium atom, for which $1/\gamma_0=26.24$ ns \cite{steck2001rubidium}. 
	%It was observed in the theoretical calculations that the TNF radius is a major factor affecting the decay rate of the atoms. By considering an uncertainty of 10\% in the TNF radius, the theoretical lifetime evaluated is $ 24.42 \pm 1.27$ ns which agrees with the experimental result . By using the cooling laser for the MOT as the Local Oscillator in the detector, the number of acquisitions required for a good fit can be reduced. Also the accumulation of Rubidium on the TNF surface causes loss of transmission and hence a reduction in the signal intensity. A possible solution is to employ a moderately powerful probe beam ($\approx 500\mu W$) to heat the atoms from the surface \cite{Pittman}.
	%In conclusion, we have demonstrated a method to evaluate the emission lifetime of Rubidium-87 atoms in the presence of a tapered nanofiber. This method can act as a tool to characterize the interactions between atoms and the TNF.
	
	\section{Theoretical model}
	We ascribe this discrepancy to two factors. First, the field we observe is resonant fluorescence rather than spontaneous emission: the atom is being excited at the same time as it is emitting. The spectrum of the resonance fluorescence is described by the central peak of the Mollow triplet, whose width depends on the Rabi frequency of the pump field as well as its detuning from the resonance. In our case, we estimate the Rabi frequency of the MOT trapping beams as $2.2\gamma$ and their detuning as $-2.6\gamma$, which yields the peak width of $1.6\gamma_0$  \cite{AgarwalBook}. 
	
\begin{figure}[h]
		\centering
		\includegraphics[width=0.75\linewidth]{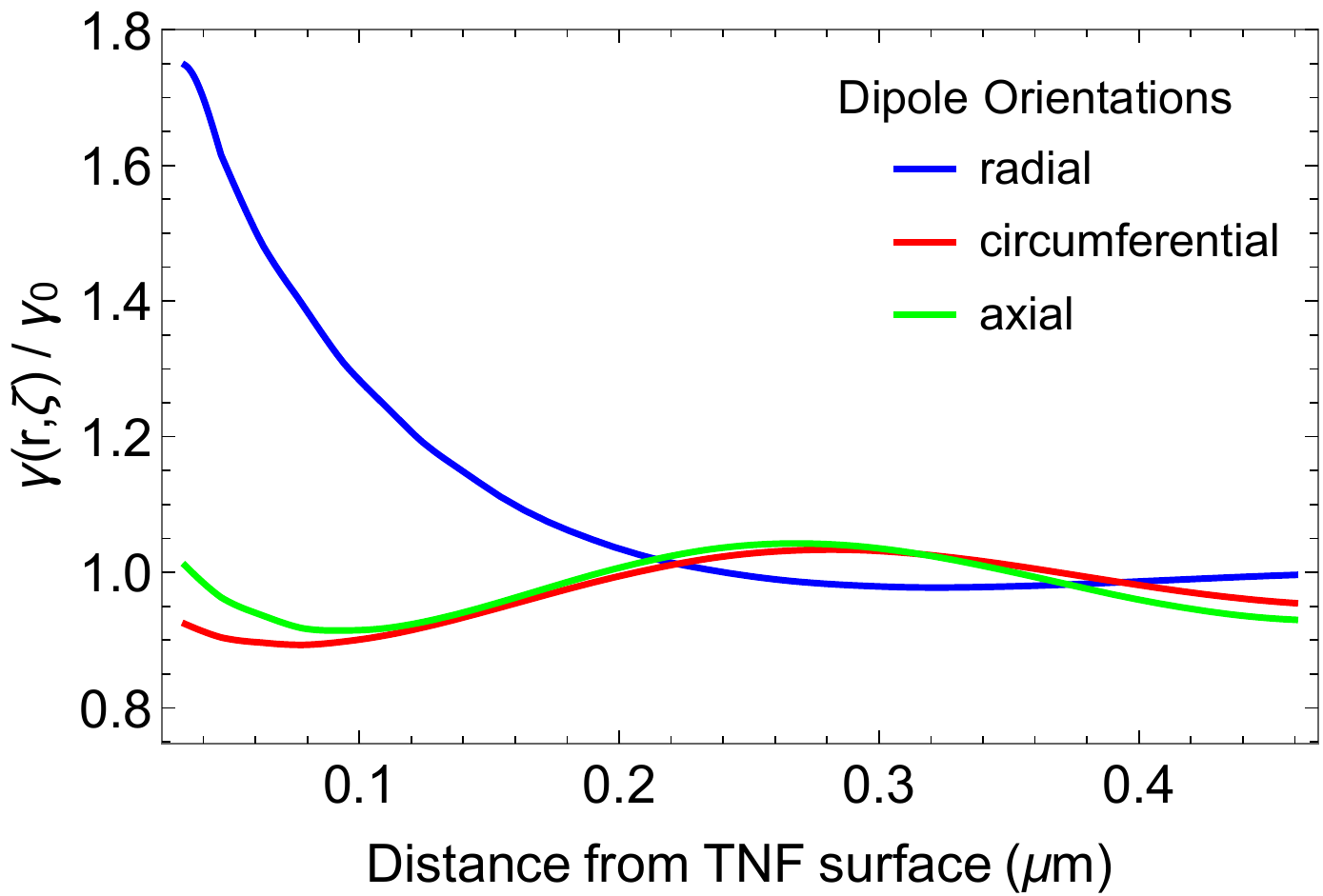}
		\caption{Theoretically calculated ratio of the electromagnetic power produced by a classical emitter in the neighborhood of a TNF to that without TNF as a function of the atomic distance from the fiber surface. The TNF has a radius of 230 nm. The three plots correspond to different orientations of the dipole with respect to the TNF.}
		\label{fig:flux}
	\end{figure}

	Second, the TNF modifies the electromagnetic mode structure around an atom, and therefore alters the atomic spontaneous emission rate due to Purcell-like effects. The theoretical analysis of this modification is done by considering a classical oscillating dipole in the vicinity of the TNF. The spontaneous decay rate of an emitter is directly proportional to the power emitted by such a classical dipole \cite{dipole}. We evaluate this power by finite-difference time-domain simulation using the software package MEEP \cite{meep}. The simulation uses a monochromatic point dipole source in a 3D simulation lattice of size $20\times20\times20$ wavelengths with the nanofiber passing through its center. A one-wavelength thick perfectly matched layer boundary is used to model the decay of the fields at the edges of the simulation volume. The electromagnetic field energy flux through this boundary is evaluated for different positions and orientations of the source. The Purcell enhancement is determined by taking the ratio of that flux in the presence and absence of the TNF. It exhibits dependence on the orientation of the emitter and its distance from the TNF as seen in Fig.~\ref{fig:flux}. The effect fades  within $\sim400$ nm from the fiber surface.

	Since our system contains many atoms, the autocorrelation function of the BHD photocurrent is obtained by averaging all possible positions and orientations, so Eq.~(\ref{rhospont}) takes the form 
	\begin{equation}\label{rhospontint}
	\rho_{jk}=\frac13\theta(j)\theta(k)\sum\limits_{\zeta=1}^3\int\gamma(r,\zeta)\alpha(r,\zeta)\tau e^{-\gamma(r,\zeta)\tau(j+k)/2}n(r)\de V,
	\end{equation}
	where $\gamma(r,\zeta)$ is the net decay rate as a function of the dipole orientation $\zeta$ and its distance $r$ from the fiber surface (Fig.~3), $n(r)$ is the atomic number density, which we assume spatially independent and $\alpha(r)$ is a weight factor which defines the fraction of the emission that goes into the guided mode. The coupling into the guided mode depends on the dipole's orientation and its distance from the surface \cite{masalov}. Mathematically, it is given by
	\begin{equation}\label{eq:2}
	\alpha(r)=\frac{\gamma(r,\zeta)-\gamma_0}{\gamma(r,\zeta)}
	\end{equation}
	Since the atoms closer to the TNF give a stronger contribution to the intensity, we expect a rapidly decaying weight factor. Averaging the density matrix according to Eq.~(\ref{rhospontint}), we obtain a decay profile that is very well approximated by an exponential with the decay rate $\gamma_{\rm avg}\approx(1.25\pm0.1)\gamma_0$, where the uncertainty is due to that in the fiber radius. 
	
	The aggregate action of the two factors modifying the decay rate gives a factor of $1.6\times 1.25=2.0\pm 0.15$, consistent with our experimental observation.
	
	Our experimental data permit us to estimate the mean number of photons emitted into the nanofiber by the surrounding atoms. The average quadrature variance associated with the single-photon state is three times that of the vacuum. That is, if each time interval $\tau$ in the measured mode contained a photon, we would observe $A_V(0)-A_{V0}(0)=2A_{V0}(0)$. In fact, we observe $A_V(0)-A_{V0}(0)=2.5\times 10^{-4}A_{V0}(0)$. This means the photons arrive at an average rate of  $r=1.25\times10^{-4}\tau^{-1}=6.25\times 10^4$ s$^{-1}$. This is consistent with our estimate made with the SPCM, given that the BHD is sensitive to only one polarization mode.
	
	The MOT has a density of about $10^{10}-10^{11}$ atoms/cm$^{3}$ resulting in around $30-300$ atoms within a radius of $500$ nm around the 3-mm TNF waist. The average weight factor $\alpha (r,\zeta)$ integrated over this neighbourhood of the TNF is about $10^{-3}$. Estimating about a half of the population to be in the excited state at any given time, we expect the photon emission rate into the fiber mode to be on a scale of $50-500$ KHz.  This is consistent with the measured photon count rate.
	
	The use of evanescent dipole traps \cite{Vetsch2010} would enhance the average atom number by 3--4 orders of magnitude, making the contribution of the atomic fluorescence to the homodyne signal comparable to the shot noise. Under these circumstances, not only the temporal properties, but also complete quantum state characterization of the spontaneously emitted light would become possible to measure. As a final comment, the fact that atomic fluorescence can be measured with balanced homoodyne detection may appear paradoxical because fluorescence is seen as an inherently incoherent phenomenon whereas nonzero autocorrelation of the BHD output voltage indicates the presence of some degree of coherence in that emission. Of course, this coherence is nothing but a consequence of a finite spectral width of the fluorescence \cite{Matthiesen2012}.

	\section{Funding Information}
	
	This project was supported by NSERC. AL is a CIFAR Fellow.
	
	% Bibliography
	\bibliography{SpontNanofiber}
	
	% Full bibliography added automatically for Optics Letters submissions; the following line will simply be ignored if submitting to other journals.
	% Note that this extra page will not count against page length
	\bibliographyfullrefs{SpontNanofiber}

\end{document}